\documentclass[aps,prl,reprint,groupedaddress]{revtex4-1}
\usepackage{graphicx} 
\usepackage{dcolumn} 
\usepackage{bm}
\usepackage{hyperref}

\begin{document}

\title{Production of new superheavy Z=108-114 nuclei with $^{238}$U, $^{244}$Pu and $^{248,250}$Cm targets}

\author{Zhao-Qing Feng}
\email{fengzhq@impcas.ac.cn}
\author{Gen-Ming Jin}
\author{Jun-Qing Li}
\affiliation{Institute of Modern Physics, Chinese Academy of
Sciences, Lanzhou 730000, People's Republic of China}

\date{\today}

\begin{abstract}
Within the framework of the dinuclear system (DNS) model, production
cross sections of new superheavy nuclei with charged numbers
Z=108-114 are analyzed systematically. Possible combinations based
on the actinide nuclides $^{238}$U, $^{244}$Pu and $^{248,250}$Cm
with the optimal excitation energies and evaporation channels are
pointed out to synthesize new isotopes which lie between the
nuclides produced in the cold fusion and the $^{48}$Ca induced
fusion reactions experimentally, which are feasible to be
constructed experimentally. It is found that the production cross
sections of superheavy nuclei decrease drastically with the charged
numbers of compound nuclei. Larger mass asymmetries of the entrance
channels enhance the cross sections in 2n-5n channels.
\begin{description}
\item[PACS number(s)]
25.70.Jj, 24.10.-i, 25.60.Pj
\end{description}
\end{abstract}

\maketitle

The synthesis of superheavy nuclei (SHN) is motivated with respect
to search the "island of stability" which is predicted
theoretically, and has obtained much experimental research with
fusion-evaporation reactions \cite{Ho00,Og07}. Neutron-deficient SHN
with charged numbers Z=107-112 were synthesized in cold fusion
reactions with the $^{208}$Pb and $^{209}$Bi targets for the first
time and investigated at GSI (Darmstadt, Germany) with the heavy-ion
accelerator UNILAC and the SHIP separator \cite{Ho00,Mu99}.
Recently, experiments on the synthesis of element 113 in the
$^{70}$Zn+$^{209}$Bi reaction have been performed successfully at
RIKEN (Tokyo, Japan) \cite{Mo04}. More neutron-rich SHN with
Z=113-116, 118 were assigned at FLNR in Dubna (Russia) with the
double magic nucleus $^{48}$Ca bombarding the actinide nuclei
\cite{Og07,Og04,Og06}. New heavy isotopes $^{259}$Db and $^{265}$Bh
were also synthesized at HIRFL in Lanzhou (China) \cite{Ga01}. New
SHN between the isotopes of the cold fusion and the $^{48}$Ca
induced reactions are of importance not only for investigating the
structure of SHN such as influence of shell effect on stability of
SHN etc, and also as a stepstone for further synthesizing and
identifying heavier superheavy nuclei.

Here we use a dinuclear system (DNS) model \cite{Fe06,Fe07}, in
which the nucleon transfer is coupled to the relative motion by
solving a set of microscopically derived master equations by
distinguishing protons and neutrons, and a barrier distribution in
the capture and fusion process of two colliding nuclei is introduced
in the model. In the DNS model, the evaporation residue cross
section is expressed as a sum over partial waves with angular
momentum $J$ at centre-of-mass energy $E_{c.m.}$ \cite{Fe07,Fe09},
\begin{eqnarray}
\sigma_{ER}(E_{c.m.})=&&\frac{\pi \hbar^{2}}{2\mu
E_{c.m.}}\sum_{J=0}^{J_{max}}(2J+1) T(E_{c.m.},J) \nonumber\\
&&\times P_{CN}(E_{c.m.},J)W_{sur}(E_{c.m.},J).
\end{eqnarray}
Here, $T(E_{c.m.},J)$ is the transmission probability of the two
colliding nuclei overcoming the Coulomb barrier in the entrance
channel to form the DNS. The $P_{CN}$ is the probability that the
system will evolve from a touching configuration to the formation of
compound nucleus in competition with the quasi-fission of the DNS
and the fission of heavy fragment. The last term is the survival
probability of the formed compound nucleus, which can be estimated
with the statistical evaporation model by considering the
competition between neutron evaporation and fission \cite{Fe06}.

Within the concept of the DNS, the fusion probability was also
calculated by using the multidimensional Kramers-type expression to
get the fusion and quasifission rate by Adamian \emph{et al.}
\cite{Ad97}. In order to describe the fusion dynamics as a diffusion
process along proton and neutron degrees of freedom, the fusion
probability is obtained by solving a set of master equations
numerically in the potential energy surface of the DNS. The time
evolution of the distribution probability function
$P(Z_{1},N_{1},E_{1},t)$ for fragment 1 with proton number $Z_{1}$
and neutron number $N_{1}$ and with excitation energy $E_{1}$ is
described by the following master equations,
\begin{widetext}
\begin{eqnarray}
\frac{d P(Z_{1},N_{1},E_{1},t)}{dt}=&&\sum_{Z_{1}^{\prime
}}W_{Z_{1},N_{1};Z_{1}^{\prime},N_{1}}(t)\left[
d_{Z_{1},N_{1}}P(Z_{1}^{\prime},N_{1},E_{1}^{\prime},t)-d_{Z_{1}^{\prime
},N_{1}}P(Z_{1},N_{1},E_{1},t)\right]+\sum_{N_{1}^{\prime
}}W_{Z_{1},N_{1};Z_{1},N_{1}^{\prime}}(t) \nonumber \\
&& \left[
d_{Z_{1},N_{1}}P(Z_{1},N_{1}^{\prime},E_{1}^{\prime},t)-d_{Z_{1},N_{1}^{\prime}}P(Z_{1},N_{1},E_{1},t)\right]-
\left[\Lambda_{qf}(\Theta(t))+\Lambda_{fis}(\Theta(t))
\right]P(Z_{1},N_{1},E_{1},t).
\end{eqnarray}
\end{widetext}
Here $W_{Z_{1},N_{1};Z_{1}^{\prime},N_{1}}$
($W_{Z_{1},N_{1};Z_{1},N_{1}^{\prime}}$) is the mean transition
probability from the channel $(Z_{1},N_{1},E_{1})$ to
$(Z_{1}^{\prime},N_{1},E_{1}^{\prime})$ (or $(Z_{1},N_{1},E_{1})$ to
$(Z_{1},N_{1}^{\prime},E_{1}^{\prime})$), and $d_{Z_{1},N_{1}}$
denotes the microscopic dimension corresponding to the macroscopic
state $(Z_{1},N_{1},E_{1})$. The sum is taken over all possible
proton and neutron numbers that fragment
$Z_{1}^{\prime},N_{1}^{\prime}$ may take, but only one nucleon
transfer is considered in the model with the relation $Z_{1}^{\prime
}=Z_{1}\pm 1$ and $N_{1}^{\prime }=N_{1}\pm 1$. The excitation
energy $E_{1}$ is determined by the dissipation energy from the
relative motion and the potential energy surface of the DNS. The
motion of nucleons in the interacting potential is governed by the
single-particle Hamiltonian \cite{Fe06,Fe07}. The evolution of the
DNS along the variable R leads to the quasi-fission of the DNS. The
quasi-fission rate $\Lambda_{qf}$ and the fission rate
$\Lambda_{fis}$ of heavy fragment are estimated with the
one-dimensional Kramers formula \cite{Fe07,Fe09}.

In the relaxation process of the relative motion, the DNS will be
excited by the dissipation of the relative kinetic energy. The local
excitation energy is determined by the excitation energy of the
composite system and the potential energy surface of the DNS. The
potential energy surface (PES) of the DNS is given by
\begin{eqnarray}
U(\{\alpha\})=&&B(Z_{1},N_{1})+B(Z_{2},N_{2})-
\left[B(Z,N)+V^{CN}_{rot}(J)\right] \nonumber\\
&&+V(\{\alpha\})
\end{eqnarray}
with $Z_{1}+Z_{2}=Z$ and $N_{1}+N_{2}=N$. Here the symbol
$\{\alpha\}$ denotes the sign of the quantities $Z_{1}, N_{1},
Z_{2}, N_{2}; J, \textbf{R}; \beta_{1}, \beta_{2}, \theta_{1},
\theta_{2}$. The $B(Z_{i},N_{i}) (i=1,2)$ and $B(Z,N)$ are the
negative binding energies of the fragment $(Z_{i},N_{i})$ and the
compound nucleus $(Z,N)$, respectively, which are calculated from
the liquid drop model, in which the shell and the pairing
corrections are included reasonably. The $V^{CN}_{rot}$ is the
rotation energy of the compound nucleus. The $\beta_{i}$ represent
the quadrupole deformations of the two fragments. The $\theta_{i}$
denote the angles between the collision orientations and the
symmetry axes of deformed nuclei. The interaction potential between
fragment $(Z_{1},N_{1})$ and $(Z_{2},N_{2})$ includes the nuclear,
Coulomb and centrifugal parts, the details are given in Ref.
\cite{Fe07}. In the calculation, the distance $\textbf{R}$ between
the centers of the two fragments is chosen to be the value which
gives the minimum of the interaction potential, in which the DNS is
considered to be formed. So the PES depends on the proton and
neutron numbers of the fragments.

The formation probability of the compound nucleus at the Coulomb
barrier $B$ and for the angular momentum $J$ is given by
\cite{Fe06,Fe07}
\begin{equation}
P_{CN}(E_{c.m.},J,B)=\sum_{Z_{1}=1}^{Z_{BG}}\sum_{N_{1}=1}^{N_{BG}}P(Z_{1},N_{1},E_{1},\tau
_{int}).
\end{equation}
The interaction time $\tau _{int}$ in the dissipation process of two
colliding partners is dependent on the incident energy $E_{c.m.}$
and the quantities $J$ and $B$. We obtain the fusion probability as
\begin{equation}
P_{CN}(E_{c.m.},J)=\int f(B)P_{CN}(E_{c.m.},J,B)dB,
\end{equation}
where the barrier distribution function is taken as an asymmetric
Gaussian form.

Neutron-deficient SHN with charged numbers Z=107-113 were
synthesized successfully in the cold fusion reactions. The
evaporation residues was observed by the consecutive $\alpha$ decay
until to take place the spontaneous fission of the known nuclides,
in which the fusion dynamics and the structure properties of the
compound nucleus have a strong influence in the production of SHN.
Recently more neutron-rich and heavier SHN with charged numbers
Z=113-116, 118 were produced in the fusion-evaporation reactions of
$^{48}$Ca bombarding actinide targets. Superheavy residues were also
identified by the consecutive $\alpha$ decay, unfortunately to
spontaneous fission of unknown nuclides. Neutron-rich
projectile-target combinations are necessary to be chosen so that
superheavy residues approach the "island of stability" with the
doubly magic shell closure beyond $^{208}$Pb at the position of
protons Z=114-126 and neutrons N=184. New SHN between the isotopes
of the cold fusion and the $^{48}$Ca induced reactions are of
importance for the structure studies themselves and also as daughter
nuclides for identifying heavier SHN in the future.

\begin{figure*}
\includegraphics{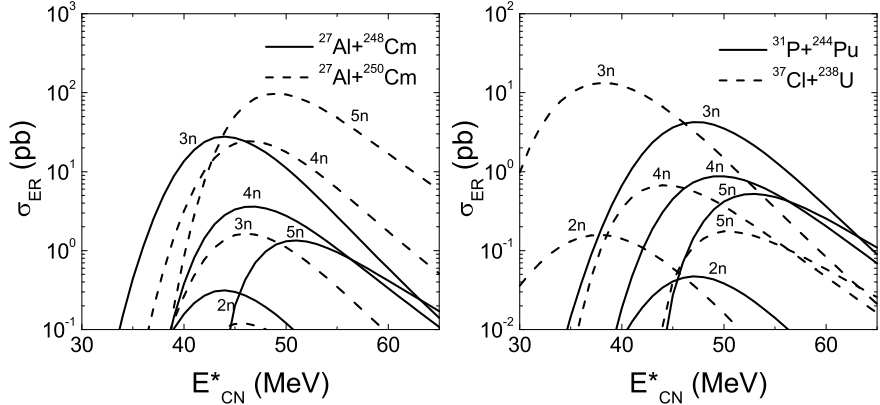}
\caption{\label{fig:wide} Evaporation residue excitation functions
in the production of isotopes of superheavy element Mt in the
reactions $^{27}$Al+$^{248,250}$Cm, $^{31}$P+$^{244}$Pu and
$^{37}$Cl+$^{238}$U.}
\end{figure*}

\begin{figure*}
\includegraphics{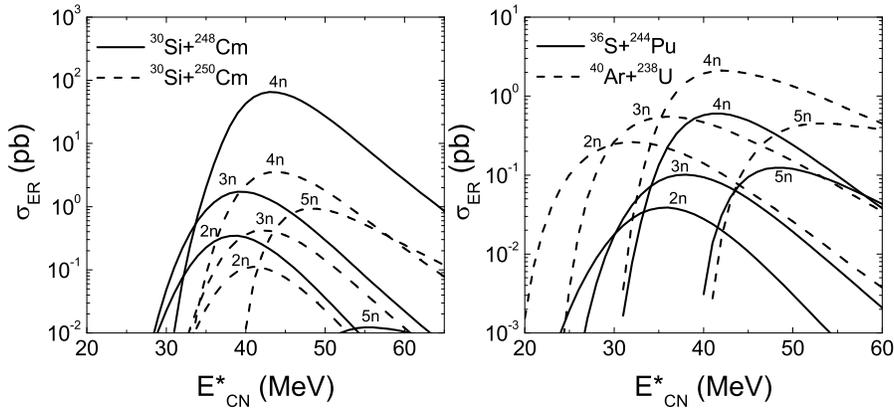}
\caption{\label{fig:wide} Calculated production cross sections for
the reactions $^{30}$Si+$^{248,250}$Cm, $^{36}$S+$^{244}$Pu and
$^{40}$Ar+$^{238}$U to produce superheavy element Ds.}
\end{figure*}

The excitation energy of compound nucleus is obtained by
$E^{\ast}_{CN}=E_{c.m.}+Q$, where $E_{c.m.}$ is the incident energy
in the center-of-mass system. The $Q$ value is given by $Q=\Delta
M_{P}+\Delta M_{T}-\Delta M_{C}$, and the corresponding mass defects
are taken from Ref. \cite{Mo95} for projectile, target and compound
nucleus, respectively. Usually, neutron-rich projectiles are used to
synthesize SHN experimentally, such as $^{64}$Ni and $^{70}$Zn in
the cold fusion reactions, which can enhance the survival
probability $W_{sur}$ in Eq.(1) of the excited compound nucleus
owing to smaller neutron separation energy. Within the framework of
the DNS model, we calculated the evaporation residue cross sections
of superheavy element Mt based on the actinide targets
$^{248,250}$Cm, $^{244}$Pu and $^{238}$U with the neutron-rich
projectiles $^{27}$Al, $^{31}$P and $^{37}$Cl as shown in Fig. 1.
One can see that the 3n channel in the reactions
$^{27}$Al+$^{248}$Cm and $^{37}$Cl+$^{238}$U, and the 4n and 5n
channels in the system $^{27}$Al+$^{250}$Cm have the larger cross
sections in the production of SHN $^{272}$Mt and $^{273}$Mt.
Superheavy element Ds(Z=110) was successfully synthesized in the
cold fusion reactions \cite{Ho00,Ho95}. The production of the SHN
depends on the isotopic combinations of projectiles and targets.
Calculations were performed for the reactions
$^{30}$Si+$^{248,250}$Cm, $^{36}$S+$^{244}$Pu and
$^{40}$Ar+$^{238}$U to produce superheavy element Ds as shown in
Fig. 2. Combination with $^{248}$Cm has the larger cross section in
the 4n channel than the isotope $^{250}$Cm due to the larger value
of survival probability. The 4n channels in the systems
$^{30}$Si+$^{248,250}$Cm and $^{40}$Ar+$^{238}$U and the 3n channel
in the reaction $^{30}$Si+$^{248}$Cm are feasible in the synthesis
of new SHN $^{274-276}$Ds. These combinations can be chosen in
experimental preparation with the present facilities.

\begin{figure*}
\includegraphics{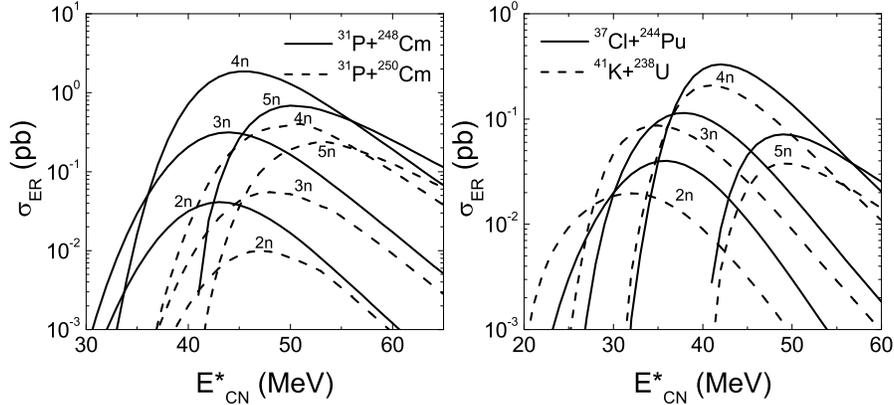}
\caption{\label{fig:wide} The same as in Fig. 1, but for the
reactions $^{31}$P+$^{248,250}$Cm, $^{37}$Cl+$^{244}$Pu and
$^{41}$K+$^{238}$U to produce superheavy element Rg.}
\end{figure*}

\begin{figure*}
\includegraphics{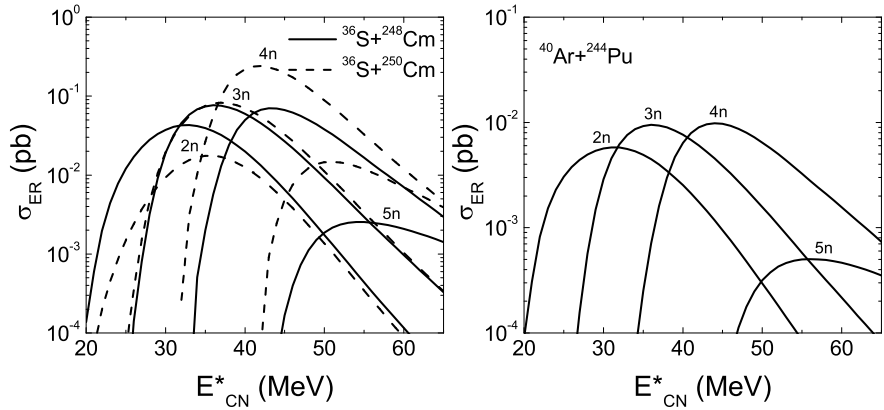}
\caption{\label{fig:wide} Comparison of the calculated production
cross sections in the reactions $^{36}$S+$^{248,250}$Cm and
$^{40}$Ar+$^{244}$Pu to synthesize superheavy element Z=112.}
\end{figure*}

In the DNS model, the isotopic trends are mainly determined by both
the fusion and survival probabilities. When the neutron number of
the projectile is increasing, the DNS gets more symmetrical and the
fusion probability decreases if the DNS does not consist of more
stable nuclei due to a higher inner fusion barrier. A smaller
neutron separation energy and a larger shell correction lead to a
larger survival probability. The compound nucleus with closed
neutron shells has larger shell correction energy and neutron
separation energy. The cross section decreases rapidly in the
production of the isotopes of Rg(Z=111). Optical channels are the 4n
evaporation for the systems $^{31}$P+$^{248,250}$Cm,
$^{37}$Cl+$^{244}$Pu and $^{41}$K+$^{238}$U to produce
$^{275,277}$Rg as shown in Fig. 3. Superheavy element Z=112 is more
difficult to be produced in the selected systems. Shown in Fig. 5
gives that the possible way is the 4n channel in the reaction
$^{36}$S+$^{250}$Cm, but the cross section is still smaller than the
system $^{48}$Ca+$^{238}$U \cite{Fe09} although the larger mass
asymmetry.

\begin{table*}
\caption{\label{tab:table} Comparisons of calculated maximal
evaporation residue cross sections and optimal excitation energies
(in bracket) in 2n-5n channels.}
\begin{ruledtabular}
\begin{tabular}{cccccc}

&Reactions &$\sigma_{ER}^{2n}$(pb) ($E^{\ast}_{CN}$) &$\sigma_{ER}^{3n}$(pb) ($E^{\ast}_{CN}$) &$\sigma_{ER}^{4n}$(pb) ($E^{\ast}_{CN}$) &$\sigma_{ER}^{5n}$(pb) ($E^{\ast}_{CN}$)\\
\hline
&$^{26}$Mg+$^{248}$Cm  &2.50 (40 MeV)  &26.2  (40 MeV)  &719.1 (42 MeV)  &1.23  (51 MeV) \\
&$^{26}$Mg+$^{250}$Cm  &1.11 (41 MeV)  &10.57 (41 MeV)  &185.2 (42 MeV)  &108.5 (45 MeV) \\
&$^{30}$Si+$^{244}$Pu  &0.46 (42 MeV)  &5.09  (43 MeV)  &185.1 (44 MeV)  &0.72  (51 MeV) \\
&$^{36}$S+$^{238}$U    &0.21 (37 MeV)  &1.96  (38 MeV)  &42.97 (42 MeV)  &0.11  (52 MeV) \\
&$^{27}$Al+$^{248}$Cm  &0.31 (44 MeV)  &27.83 (44 MeV)  &3.59  (47 MeV)  &1.34  (51 MeV) \\
&$^{27}$Al+$^{250}$Cm  &0.12 (46 MeV)  &1.64  (46 MeV)  &24.31 (46 MeV)  &97.44 (49 MeV) \\
&$^{31}$P+$^{244}$Pu   &4.71$\times10^{-2}$ (47 MeV)    &4.25 (47 MeV)   &0.87 (50 MeV)  &0.52 (53 MeV) \\
&$^{37}$Cl+$^{238}$U   &0.16 (38 MeV)  &13.31 (38 MeV)  &0.67 (44 MeV)   &0.17  (50 MeV) \\
&$^{30}$Si+$^{248}$Cm  &0.34 (39 MeV)  &1.72  (39 MeV)  &65.32 (43 MeV)  &1.22$\times10^{-2}$  (56 MeV) \\
&$^{30}$Si+$^{250}$Cm  &0.11 (41 MeV)  &0.42  (42 MeV)  &3.54 (44 MeV)   &0.93 (48 MeV) \\
&$^{36}$S+$^{244}$Pu   &3.87$\times10^{-2}$ (36 MeV)    &0.101 (38 MeV)  &0.61 (41 MeV)  &0.12 (48 MeV) \\
&$^{40}$Ar+$^{238}$U   &0.26 (32 MeV)  &0.55 (36 MeV)   &2.10 (42 MeV)   &0.45 (53 MeV) \\
&$^{31}$P+$^{248}$Cm   &4.11$\times10^{-2}$ (43 MeV)  &0.31  (44 MeV)  &1.85 (45 MeV)  &0.69  (50 MeV) \\
&$^{31}$P+$^{250}$Cm   &9.91$\times10^{-3}$ (47 MeV)  &5.49$\times10^{-2}$  (48 MeV)  &0.41 (51 MeV)   &0.25 (52 MeV) \\
&$^{37}$Cl+$^{244}$Pu  &4.01$\times10^{-2}$ (36 MeV)  &0.11 (38 MeV)  &0.33 (42 MeV)  &7.15$\times10^{-2}$ (49 MeV) \\
&$^{41}$K+$^{238}$U    &1.96$\times10^{-2}$ (32 MeV)  &8.67$\times10^{-2}$ (35 MeV)   &0.21 (41 MeV)   &3.77$\times10^{-2}$ (49 MeV) \\
&$^{36}$S+$^{248}$Cm   &4.31$\times10^{-2}$ (33 MeV)  &7.64$\times10^{-2}$ (36 MeV)   &7.02$\times10^{-2}$ (43 MeV)  &2.55$\times10^{-3}$  (54 MeV) \\
&$^{36}$S+$^{250}$Cm   &1.76$\times10^{-2}$ (36 MeV)  &8.25$\times10^{-2}$ (37 MeV)   &0.24 (42 MeV)   &1.47$\times10^{-2}$ (51 MeV) \\
&$^{40}$Ar+$^{244}$Pu  &5.79$\times10^{-3}$ (31 MeV)  &9.48$\times10^{-3}$ (36 MeV)   &9.84$\times10^{-3}$ (44 MeV)  &5.02$\times10^{-4}$ (56 MeV) \\
&$^{37}$Cl+$^{248}$Cm  &5.81$\times10^{-2}$ (31 MeV)  &0.26 (35 MeV)  &0.195 (42 MeV) &1.05$\times10^{-2}$ (52 MeV) \\
&$^{37}$Cl+$^{250}$Cm  &2.08$\times10^{-2}$ (35 MeV)  &0.21 (36 MeV)  &0.594 (41 MeV) &7.06$\times10^{-2}$ (49 MeV) \\
&$^{41}$K+$^{244}$Pu   &1.11$\times10^{-2}$ (29 MeV)  &4.22$\times10^{-2}$ (34 MeV)  &3.31$\times10^{-2}$ (42 MeV)  &2.3$\times10^{-3}$ (53 MeV) \\
&$^{45}$Sc+$^{238}$U   &1.72$\times10^{-2}$ (27 MeV)  &1.99$\times10^{-2}$ (35 MeV)  &2.32$\times10^{-3}$ (45 MeV)  &1.92$\times10^{-4}$ (57 MeV) \\
&$^{40}$Ar+$^{248}$Cm  &6.98$\times10^{-3}$ (26 MeV)  &2.21$\times10^{-2}$ (33 MeV)  &3.5$\times10^{-2}$ (41 MeV)   &2.12$\times10^{-3}$ (51 MeV) \\
&$^{40}$Ar+$^{250}$Cm  &2.77$\times10^{-3}$ (29 MeV)  &2.11$\times10^{-2}$ (33 MeV)  &7.96$\times10^{-2}$ (40 MeV)  &9.69$\times10^{-3}$ (48 MeV) \\
&$^{48}$Ti+$^{238}$U   &4.51$\times10^{-2}$ (24 MeV)  &1.37$\times10^{-2}$ (32 MeV)  &5.81$\times10^{-3}$ (42 MeV)  &1.71$\times10^{-4}$ (54 MeV) \\
&$^{50}$Ti+$^{238}$U   &5.11$\times10^{-2}$ (23 MeV)  &2.18$\times10^{-2}$ (31 MeV)  &1.07$\times10^{-2}$ (40 MeV)  &4.11$\times10^{-4}$ (50 MeV) \\

\end{tabular}
\end{ruledtabular}
\end{table*}

The productions of superheavy element Z=113 were successfully
performed in the cold fusion reaction $^{70}$Zn+$^{209}$Bi
\cite{Mo04} and also in the hot fusion $^{48}$Ca+$^{237}$Np
\cite{Og07b} with the cross section less than 1 pb. We calculated
the evaporation residue excitation functions for the reactions
$^{37}$Cl+$^{248,250}$Cm, $^{41}$K+$^{244}$Pu and
$^{45}$Sc+$^{238}$U. The results show that the 3n and 4n channels in
the systems $^{37}$Cl+$^{248,250}$Cm have larger cross sections and
are possible to synthesize new isotopes $^{281-284}$113 in
experimentally. Superheavy element Z=114 is difficulty to be
produced from our calculations for the selected systems because of
the smaller cross sections with less than 0.1 pb for all systems. We
list the maximal production cross sections and the corresponding
excitation energies in the brackets calculated by using the DNS
model in Table 1. These selected systems and evaporation channels
are feasible to produce new isotopes between the cold fusion and the
$^{48}$Ca induced fusion reactions.

In summary, we systematically investigated the production of
superheavy residues in the fusion-evaporation reactions using the
DNS model, in which the nucleon transfer leading to the formation of
superheavy compound nucleus is described by a set of microscopically
derived master equations distinguishing the proton and neutron
transfer that are coupled to the dissipation of relative motion
energy and angular momentum. The production of new isotopes between
the gap of the cold fusion and the $^{48}$Ca induced fusion
reactions are discussed for selected systems. Optimal evaporation
channels and excitation energies corresponding to the maximal cross
sections are stated and discussed systematically.

\begin{acknowledgments}
We would like to thank Prof. Werner Scheid for carefully reading the
manuscript. This work was supported by the National Natural Science
Foundation of China under Grant Nos. 10805061 and 10775061, the
special foundation of the president fund, the west doctoral project
of Chinese Academy of Sciences, and major state basic research
development program under Grant No. 2007CB815000.
\end{acknowledgments}

\end{document}